\documentclass{appolb}
\usepackage{array}
\usepackage{color}
\usepackage{graphicx}

\newcommand{\tpz}{${}^{3\!}P_0$}

\begin{document}
\title{\mbox{$Z_0(57)$ and $E(38)$: possible surprises in the Standard
Model\thanks
{Presented by G.~Rupp at ``Excited QCD 2020'',
Krynica-Zdr\'{o}j, Poland, 2--8 Feb.\ 2020.}}}
\author{
Eef van Beveren\address{Centro de F\'{\i}sica da UC, Departamento de
F\'{\i}sica, Universidade de Coimbra, P-3004-516, Portugal}
\\[2mm]
George Rupp\address{Centro de F\'{\i}sica e Engenharia de Materiais
Avan\c{c}ados, Instituto Superior T\'{e}cnico, Universidade de Lisboa,
P-1049-001, Portugal}
}
\maketitle
\begin{abstract}
With the reported observation of the Higgs boson at the LHC, the Standard Model
of particle physics seems to be complete now as for its particle content.
However, several experimental data at low and intermediate energies indicate
that there may be two surprises. First we propose a tentative new boson
$Z_0(57)$, with a mass of about~57 GeV, on the basis of small enhancements we
observe in several experiments, using recent data obtained at the LHC as well
as much older ones from LEP. If confirmed, we interpret this new particle as a
pseudoscalar or scalar partner of a composite $Z$ vector boson. Secondly, we
advocate the existence of a very light spinless boson $E(38)$, probably a
scalar, with a mass of 38~MeV and decaying into two photons. Theoretical
arguments and experimental signals supporting such a novel light boson will be
presented, including a recent direct experimental confirmation at the
Joint Institute for Nuclear Research in Dubna.

\end{abstract}
\section{Introduction}
After the observation of a new boson compatible with the Standard-Model (SM)
Higgs boson was reported in 2012 by the ATLAS \cite{Higgs-ATLAS} and CMS
\cite{Higgs-CMS} Collaborations, no further particle discoveries at
the LHC have been announced so far. Even more significantly, the latest
Advanced Cold Molecule Experiment (ACME) \cite{ACME-II} measuring
a possible electron electric dipole moment has ruled
out \cite{kill-SUSY} any new particles that contribute maximally to CP
violation in one-loop and two-loop diagrams for masses up to about 50 TeV
and 3 TeV, respectively, so beyond LHC detection range. Thus, both 
ATLAS \cite{ATLAS-2018-025} and CMS \cite{PLB793p320} recently
carried out alternative searches at energies that had already been covered
by the Large Electron-Positron Collider, viz.\ for diphoton resonances
(65--110~GeV) \cite{ATLAS-2018-025} and SM-like extra Higgs bosons
(70--110~GeV) \cite{PLB793p320}. At the same time, CMS did
\cite{JHEP1811p161} a search for dimuon resonances in the mass range
12--70~GeV. No new discoveries were reported in these three experiments,
though small enhancements at about 95~GeV \cite{PLB793p320} and 28~GeV
\cite{JHEP1811p161} were observed. However, in the latter CMS paper we
noticed \cite{Z057} an additional minor enhancement at roughly 57~GeV, which
together with the one at 28~GeV lends further support to our earlier
\cite{threshold,triphoton} suggestion of a new spinless boson ``$Z_0(57)$''
with a mass of about 57~GeV on the basis of prior experiments. All these
experimental signals and our interpretation will be discussed in
Sec.~\ref{Z57} below.

On the other hand, there have also been numerous searches for much lighter
particles, which might account for dark matter, discrepancies in the proton
radius, or the muon's anomalous magnetic moment. Recently, the existence of
a very light boson with a mass of about 17~MeV was claimed \cite{X17}, in
order to explain apparent anomalies in $^8$Be and $^4$He nuclear transitions.
It has even been conjectured \cite{Fifth} that such an $X17$ boson might
be the mediator of a ``protophobic fifth force''. It also has led to 
countless other theoretical speculations over the past four years. In contrast,
a recent report \cite{JINR} by an experimental team at the Joint Institute
for Nuclear Research (JINR) in Dubna on the possible direct observation of a
novel light spinless boson with a mass of about 38~MeV has received no
attention whatsoever. Several years earlier, we had proposed \cite{E38_1}
this so-called $E(38)$ particle on the basis of asymmetries in leptonic
bottomonium decays and apparent oscillations in charmonium-production data,
besides different theoretical arguments.  In Ref.~\cite{E38_2} we then showed
a clear and more direct $E(38)$ signal, viz.\ in $\gamma\gamma$ decays
published \cite{COMPASS_1} by the COMPASS Collaboration, though disputed
\cite{COMPASS_2} by COMPASS yet reaffirmed \cite{E38_3} by us. In
Sec.~\ref{E38} we shall make the case for the $E(38)$.

\section{Indications of a spinless boson with a mass of about 57 GeV}
\label{Z57}
\begin{figure}[t]
\begin{center}
\begin{tabular}{c}
\includegraphics[height=150pt]{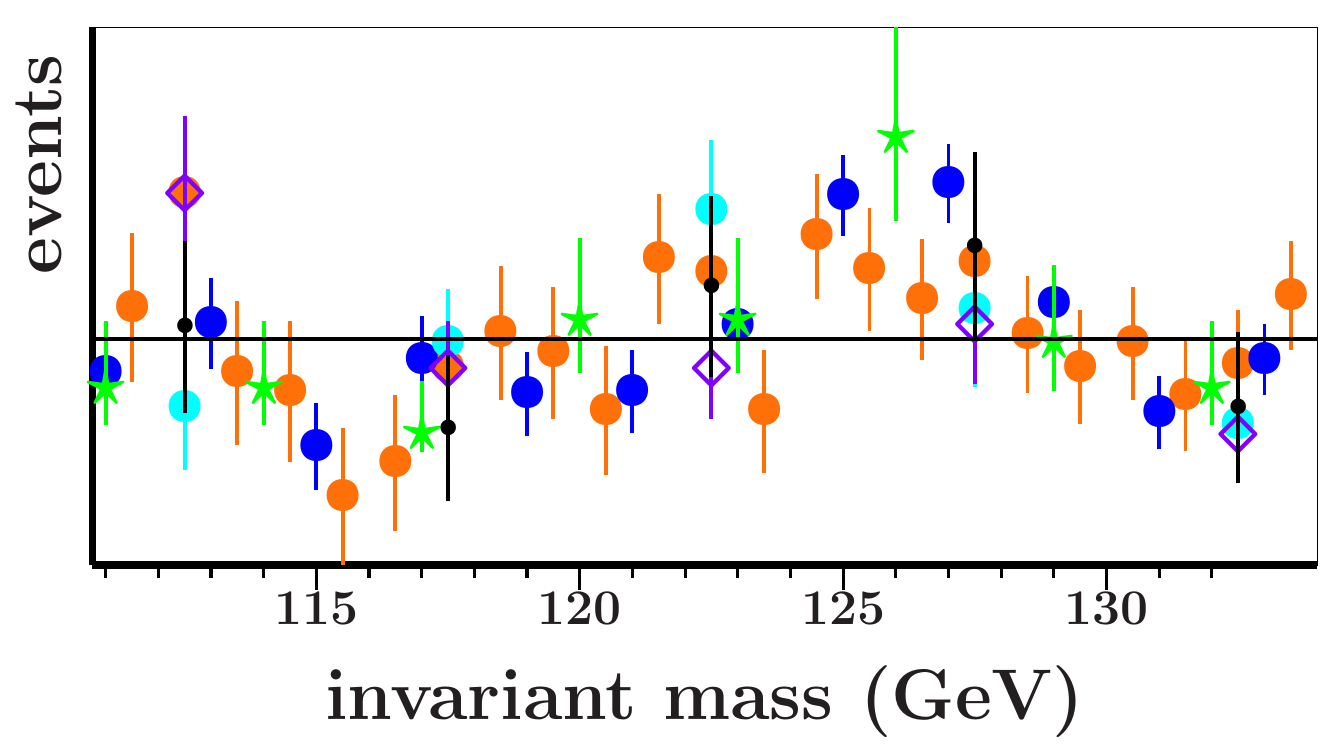}\\ [-10pt]
\end{tabular}
\end{center}
\caption[]{Diphoton
({\definecolor{tmpclr}{rgb}{1.000,0.440,0.030}{\color{tmpclr}$\bullet$}})
({\color{blue}$\bullet$}) and
four-lepton
({\color{green}$\star$})
({\definecolor{tmpclr}{rgb}{0.000,1.000,1.000}{\color{tmpclr}$\bullet$}})
signals;
invariant-mass distributions of
$\tau\tau$ in $e^{+}e^{-}\to\tau\tau (\gamma )$
({\definecolor{tmpclr}{rgb}{0.500,0.000,1.000}{\color{tmpclr}$\diamond$}}),
and
$\mu\mu$ in $e^{+}e^{-}\to\mu\mu (\gamma )$ ({$\bullet$})
(references in Refs.~\cite{threshold,Z057}). \\[-5mm]
}
\label{dip115}
\end{figure}

\begin{figure}[b]
\begin{center}
\includegraphics[width=11cm]{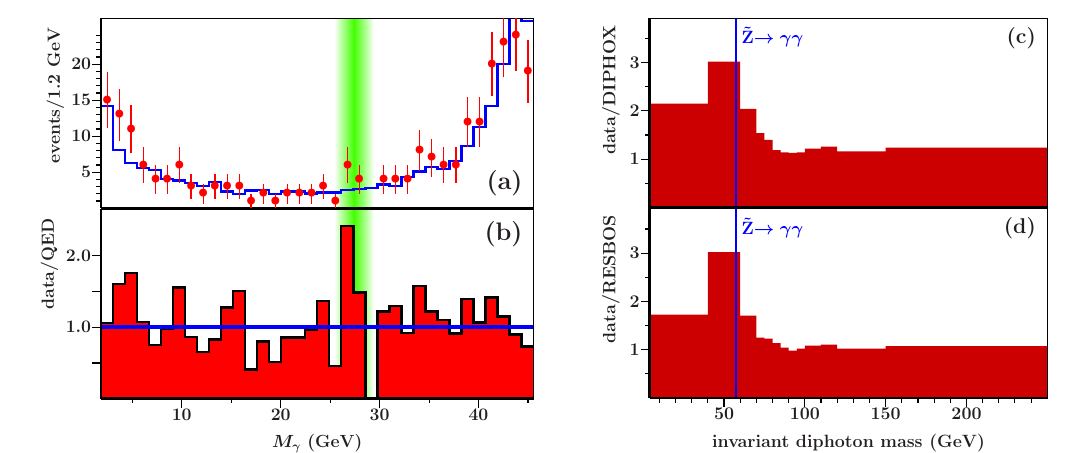} 
\end{center}
\caption[]{
(a) \cite{L3}:
Single-photon CM energies in $Z\to 3\gamma$ events
for $\sqrt{s}=M_{Z}$; histogram: Monte-Carlo expectation 
from QED; green band: expected photons from $Z\to\gamma Z_{0}$
for $M_{Z_{0}}=57.5$~GeV.
(b): As (a), but now measured/QED-expected events.
(c,d) \cite{CMS13001}: Measured/expected $\gamma\gamma$ events
for two models.
}
\label{L3CMS}
\end{figure}

\begin{figure}[!t]
\centering
\includegraphics[width=9cm]{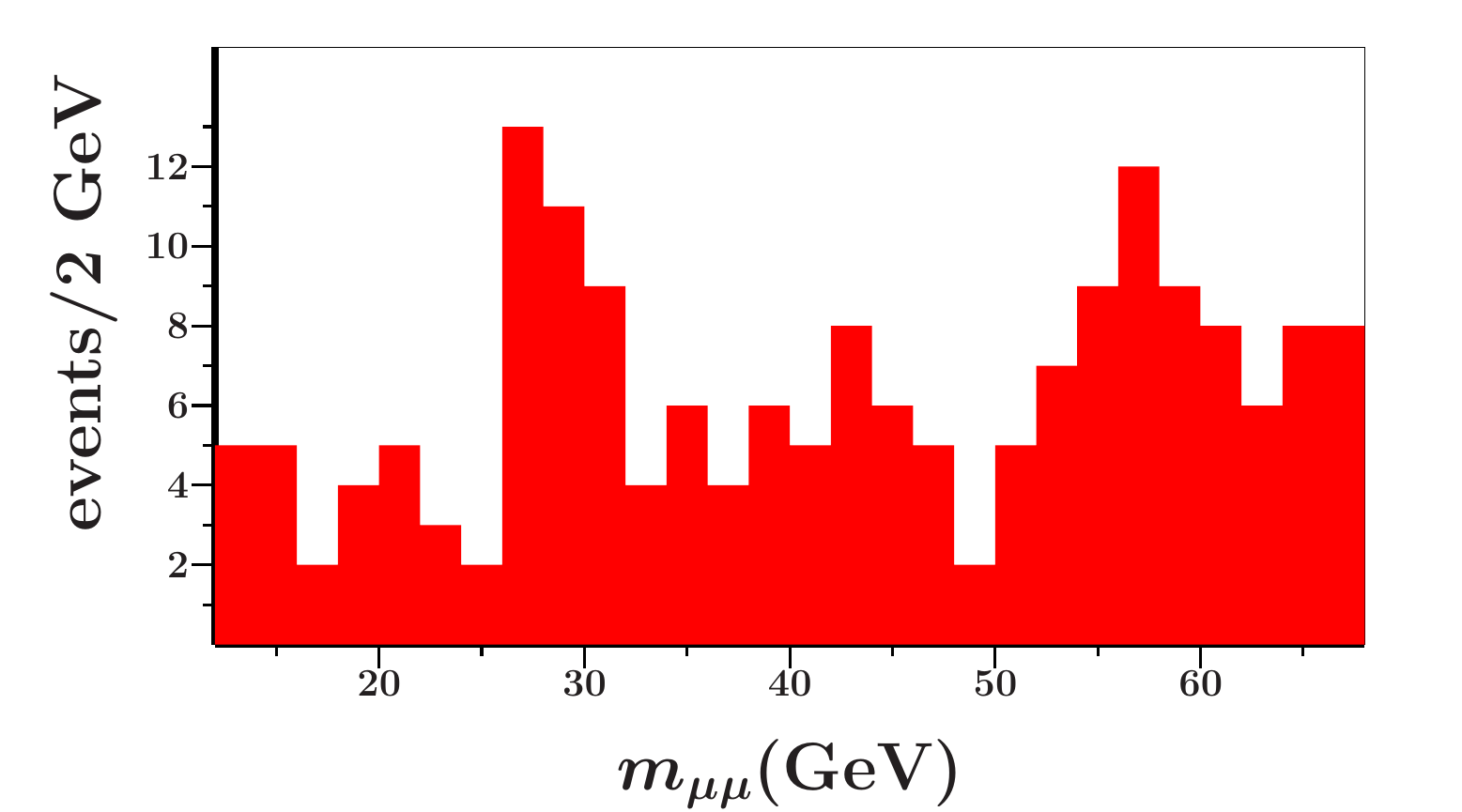} \\[-1mm]
\caption[]{\small
Data on the dimuon mass distribution in $Z$ decays,
taken from Ref.~\cite{JHEP1811p161}. \\[-5mmm]
}
\label{cms}
\end{figure}

In Ref.~\cite{threshold} we observed a conspicuous dip at about 115~GeV
in several ATLAS and CMS data, as well as in much earlier data by the
L3 Collaboration; see Fig.~\ref{dip115} and Ref.~\cite{Z057} for the
corresponding references.
The same figure reveals an enhancement slightly above 125~GeV, probably a
preliminary sign then of the Higgs boson. We interpreted \cite{threshold} the
dip as indication of a two-particle threshold opening at
115~GeV and a possible sign of a composite new boson with a mass of
about 57~GeV. Natural enhancements at two-body production thresholds of
composite particles had been predicted by us in Ref.~\cite{production}.
Possible compositeness of the $W$ and $Z$ intermediate gauge bosons has
been considered in many papers (see e.g.\ Ref.~\cite{Fritzsch}) and would
necessarily imply the existence of partner states with different quantum
numbers. This proposal received support from CMS data \cite{CMS13001}
exhibiting a modest enhancement in $\gamma\gamma$ data at about 57~GeV,
as we reported in Ref.~\cite{triphoton}, in which we also showed much
older L3 data \cite{L3} on $\gamma\gamma\gamma$ decays of the $Z$ boson;
see Fig.~\ref{L3CMS}.
In the L3 single-photon data from $Z\to\gamma\gamma\gamma$ decays one
notices a small enhancement at 28~GeV. This may be a signal of a decay
$Z\to Z_{0}\gamma$, where $Z_0$ is a new spinless boson. Moreover, the
$\gamma\gamma$ and 1$\gamma$ enhancements at 57~GeV  and 28~GeV, respectively,
perfectly satisfy the kinematics of a decay $Z\to Z_0(57)\gamma$.
Finally, additional evidence appears to come from Ref.~\cite{JHEP1811p161},
with two-muon data exhibiting an enhancement at 28~GeV and another, quite
modest one at about 57~GeV; see Fig.~\ref{cms}.
Since a photon may convert
into a muon pair in the CMS detector's strong magnetic field and a
(pseudo)scalar boson can also decay to $\mu\mu$, these recent CMS data support
our proposed $Z_0(57)$ as well, though much higher statistics will be required
for a definite confirmation.
\\[-3mm]
\section{Evidence of a very light spinless boson at 38 MeV}
\label{E38}
\begin{figure}[!bh]
\mbox{ } \\[-7mm]
\begin{center}
\begin{tabular}{c}
\includegraphics[width=340pt]{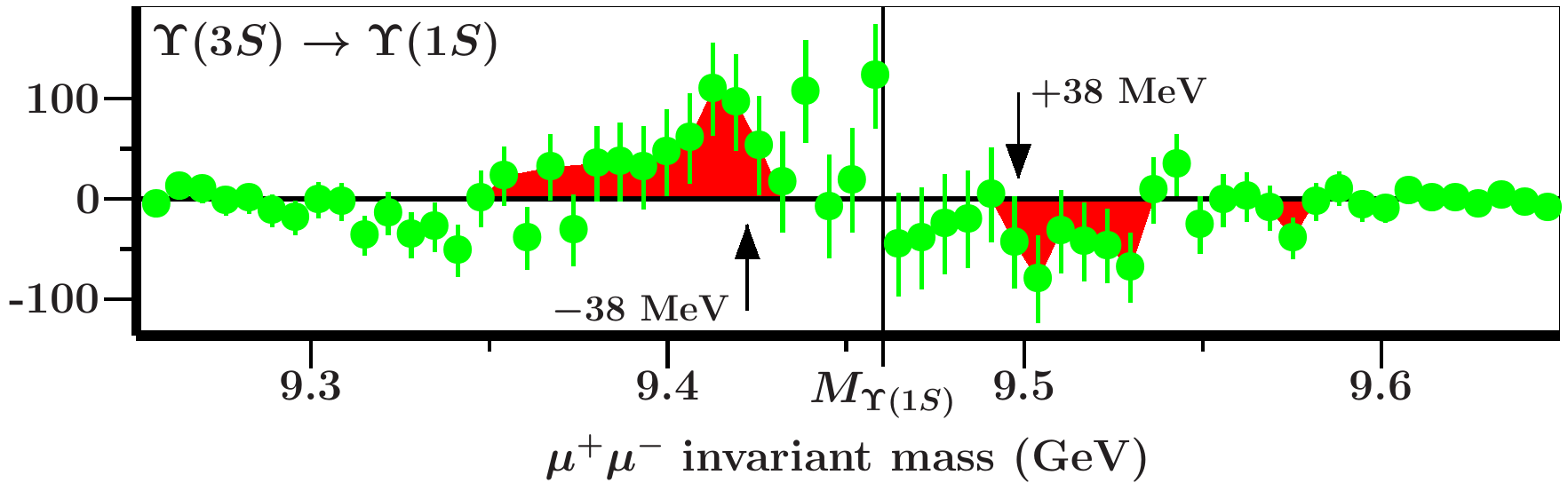}\\ [-10pt]
\end{tabular}
\end{center}
\caption{\small
Excess signal \cite{ARXIV09100423}
in the invariant $\mu^{+}\mu^{-}$ mass for the process
$\Upsilon\left( 2\,{}^{3\!}S_{1}\right)$ $\to$
$\pi^{+}\pi^{-}\Upsilon\left( 1\,{}^{3\!}S_{1}\right)$
$\to$ $\pi^{+}\pi^{-}\mu^{+}\mu^{-}$,
using bins of 6.5 MeV.
Statistical errors are shown by vertical bars.
Vertical line indicates
$M_{\mu^{+}\mu^{-}}=M_{\Upsilon\left( 1\,{}^{3\!}S_{1}\right)}$.
See further Ref.~\cite{E38_1}.
}
\label{elisa}
\end{figure}

In Ref.~\cite{E38_1} we proposed a very light new scalar boson ``$E(38)$'',
with a mass of about 38~MeV, on the basis of indirect experimental indications
and a decades-old anti-De-Sitter (AdS) model of geometric quark
confinement (see Ref.~\cite{E38_1} for references). Among the different
experimental signals, we highlight here a clear asymmetry in the $\mu^+\mu^-$
invariant mass for the decay chain
$\Upsilon\left(2\,{}^{3\!}S_{1}\right)$ $\to$
$\pi^{+}\pi^{-}\Upsilon\left(1\,{}^{3\!}S_{1}\right)$
$\to$ $\pi^{+}\pi^{-}\mu^{+}\mu^{-}$ (see Fig.~\ref{elisa}).
In Ref.~\cite{ARXIV09100423} (BABAR) both the asymmetry in the data
and the failure to explain it were acknowledged. We interpreted
\cite{E38_1} it as the undetected emission of several quanta
of a new scalar particle with a mass of about 38~MeV, showing up as minor
dips and enhancements in the asymmetric $\mu^+\mu^-$ invariant-mass
distribution, roughly at multiples of 38~MeV. Such a light scalar
could also be responsible for the empirically successful \tpz\
\cite{Micu} mechanism.

\begin{figure}[!b]
\begin{center}
\begin{tabular}{c}
\includegraphics[width=195pt]{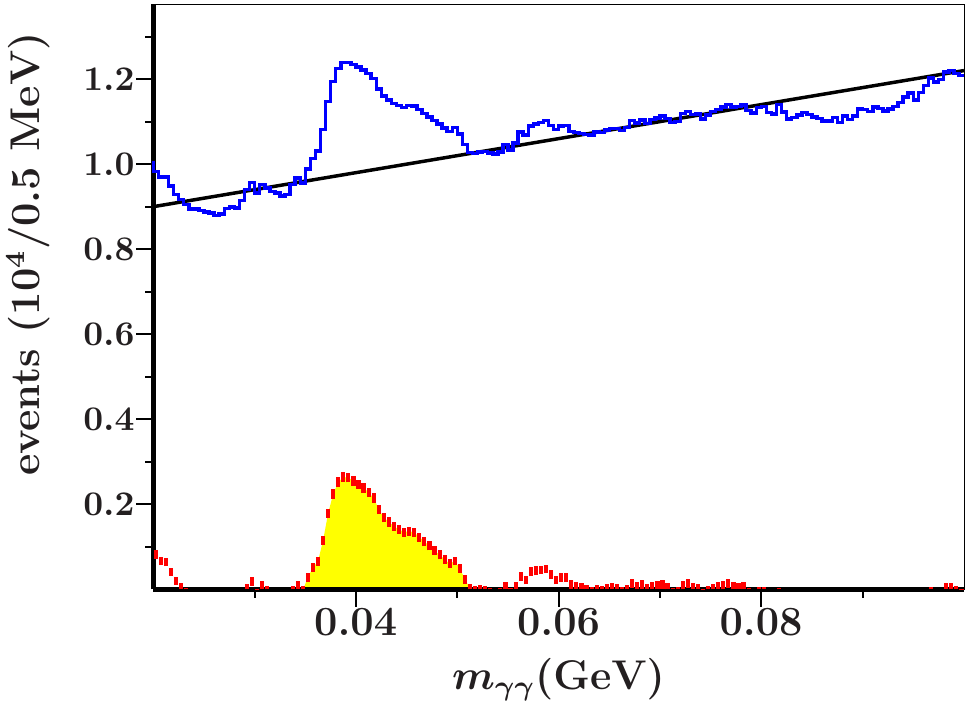} \\ [-12pt]
\end{tabular}
\end{center}
\caption{\small
Top: signal in the COMPASS \cite{COMPASS_1} $\gamma\gamma$ data,
with maximum at $\approx$39 MeV. Bottom: $E(38)$ structure
after background subtraction, with about 46,000 events.}
\label{compassgammagamma}
\end{figure}

A much more direct signal of the proposed $E(38)$ we identified \cite{E38_2}
in $\gamma\gamma$ data \cite{COMPASS_1} by the COMPASS Collaboration,
in a study of the exotic $\eta^\prime\pi^-$ wave. The corresponding
structure around 40~MeV is depicted in Fig.~\ref{compassgammagamma}. Taking
the excess data at face value, the significance of the signal is overwhelming.
However, COMPASS contested \cite{COMPASS_2} our assessment, arguing that the
bump must be an artefact resulting from secondary interactions inside the
COMPASS spectrometer as well as cuts applied to $\gamma\gamma$ events at very
low energies. To support this conclusion, a Monte-Carlo (MC) simulation was
presented in Ref.~\cite{COMPASS_2}. Nevertheless, we argued \cite{E38_3} that
the employed MC is inadequate to describe the $E(38)$-like structure and even
the actual COMPASS data below 50~MeV. In view of this controversy,
it would be good if COMPASS did a dedicated analysis at these energies.

The strongest $E(38)$ evidence was published recently \cite{JINR}
by an experimental group at the JINR in Dubna, finding significant 
$\gamma\gamma$ enhancements at 38~MeV in proton and deuteron scattering
off carbon and copper nuclei, though still lacking statistics to be
considered a particle observation. So also at JINR and other labs more 
specific experiments are highly desirable.

To conclude, a remarkable value of 38 MeV was found
for the non-perturbative contribution to the pion-nucleon $\sigma$ term,
viz.\ via a scalar light-quark tadpole \cite{QLLsM} or
condensate \cite{MPLA18p1171}. Recently, a surprising value of 38~MeV
was also obtained in a lattice computation \cite{NPPP300302p107} of the
whole $\pi N$ $\sigma$ term.

\end{document}